\documentclass[aps,prl,showpacs,showkeys,twocolumn,groupedaddress,
superscriptaddress]{revtex4}

\evensidemargin  \oddsidemargin

\usepackage{graphicx}%
\usepackage{dcolumn}

\begin{document}

\title{Quasi-Adiabatic Decay of Capillary Turbulence\\on the Charged
Surface of Liquid Hydrogen}

\author{G.~V.~Kolmakov}%
\affiliation{Institute of Solid State Physics RAS, Chernogolovka,
Moscow region, 142432, Russia}\affiliation{Department of Physics,
Lancaster University, Lancaster, LA1 4YB, UK}

\author{A.~A.~Levchenko}
\affiliation{Institute of Solid State Physics RAS, Chernogolovka,
Moscow region, 142432, Russia}

\author{M.~Yu.~Brazhnikov}
\affiliation{Institute of Solid State Physics RAS, Chernogolovka,
Moscow region, 142432, Russia}

\author{L.~P.~Mezhov-Deglin}
\affiliation{Institute of Solid State Physics RAS, Chernogolovka,
Moscow region, 142432, Russia}

\author{A.~N.~Silchenko}
\affiliation{Department of Physics, Lancaster University, Lancaster,
LA1 4YB, UK}

\author{P.~V.~E.~McClintock}
\affiliation{Department of Physics, Lancaster University, Lancaster,
LA1 4YB, UK}

\date{\today}

\begin{abstract}
We study the free decay of capillary turbulence on the charged
surface of liquid hydrogen. We find that decay begins from the
\textit{high frequency} end of the spectral range, while most of
the energy remains localized at low frequencies. The apparent
discrepancy with the self-similar theory of nonstationary wave
turbulent processes is accounted for in terms of a quasi-adiabatic
decay wherein fast nonlinear wave interactions redistribute energy
between frequency scales in the presence of finite damping at all
frequencies. Numerical calculations based on this idea agree well
with experimental data.
\end{abstract}


\maketitle

{\it Introduction.} Turbulence on the surface of fluids has been
the focus of numerous experimental and theoretical investigations
during the last few years
\cite{Book,Putterman,Alstrom,Lev1,LevEPL,Filonenko,Falkovich,Push,
Zakharov2002}. Interest in such phenomena, usually referred as
wave turbulence (WT), arises both from their great importance in
terms of basic nonlinear physics and from numerous applications in
engineering and the life sciences. One of the best known
applications is in weather prediction, using information received
from the measurements of the spectrum of waves on the sea surface.
Turbulence in a system of capillary waves is of interest because
its dynamics on this wavelength scale plays a significant role in
the transfer and dissipation of energy on the liquid surface.
Studies of nonstationary phenomena in systems of capillary waves
are of particular importance because they could provide direct
information about nonlinear wave interactions. In spite of the
large number of experimental studies devoted to the nonlinear
dynamics of surface waves, however, there are only a few recent
reports ~\cite{Putterman,Alstrom,Lev1,LevEPL} of experimental
studies of capillary turbulence that are directly comparable with
theoretical predictions.

Recently, the use of liquid hydrogen as a model medium has brought
significant progress in the understanding of capillary turbulence.
The remarkable properties of liquid hydrogen (its low viscosity
and the high nonlinearity of its capillary waves, and the
possibility of driving the electrically charged surface of the
liquid directly) have allowed us to observe for the first time:
the formation of the Kolmogorov power-law turbulent spectrum over
a wide range of frequencies (10$^2$--10$^4$~Hz)~\cite{Lev1}; the
modification of the scaling index of the turbulent spectrum in
dependence on the spectral characteristics of the driving force
\cite{ZHETF}; and the cut-off of the power spectrum of capillary
turbulence at high frequencies due to the change of the energy
transfer mechanism from nonlinear waves transformation to viscous
damping \cite{LevBound}. Many important properties of waves on the
surface of liquid hydrogen are similar to those of conventional
liquids like water (cf.\ the capillary lengths for liquid hydrogen
$\lambda=0.19$ cm at $T= 15$ K, and for water $\lambda=0.28$ cm at
$T = 293$ K), providing an additional argument for using liquid
hydrogen as a perfect test fluid for accurate tests of WT theory.
The latter are of considerable importance, because WT theory is
used to describe turbulent processes in a wide variety of media,
including e.g.\ plasmas \cite{plasma1}, astrophysics
\cite{plasma2}, ocean surfaces \cite{Newell}, acoustic turbulence
in superfluid He II \cite{HeII}, and nonlinear optics \cite{Opt}.

In this Letter we report the first observations of decay of the
turbulent state in a system of capillary waves on the surface of
liquid hydrogen. We have observed that the decay of the stationary
turbulent spectrum starts in the {\it high frequency} domain, with
the energy remaining localised in the low frequency range of the
turbulent spectrum, i.e.\ near the driving frequency. At low
frequencies the turbulent spectrum remains close to its
unperturbed shape for a relatively long period of time after the
driving force is switched off. This observation differs
significantly from what might be expected from the self-similar
theory of nonstationary WT processes \cite{Book}, where the
evolution of the spectrum is considered in the range of
frequencies where viscous damping can be neglected.

We show that where there is finite damping at all frequencies,
viscous losses cause qualitative changes in the evolution of the
turbulent spectrum after removal of the driving force: rather than
a propagation of perturbations from low to high frequencies,
caused by the cascade transfer of energy (the scenario considered
in \cite{Book}), there is a relatively fast decay of the high
frequency domain of the spectrum. Although our observations are
new, and came as a surprise, we show that they can still be
understood qualitatively within the framework of the general WT
theory \cite{Book}.


\begin{figure}[th]
\includegraphics[width=66mm]{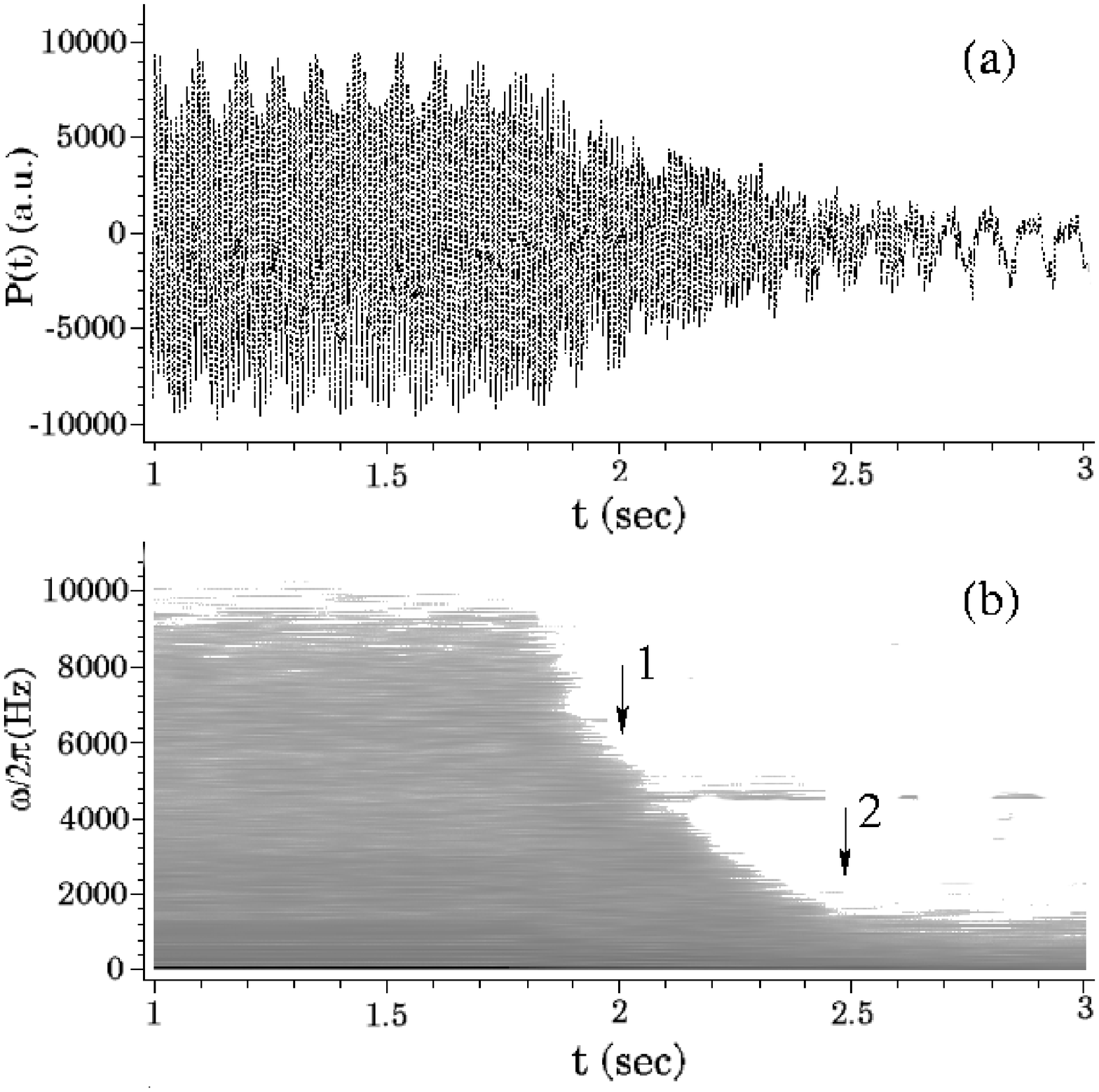}
\vskip 0.3 cm \noindent
\includegraphics[width=60mm]{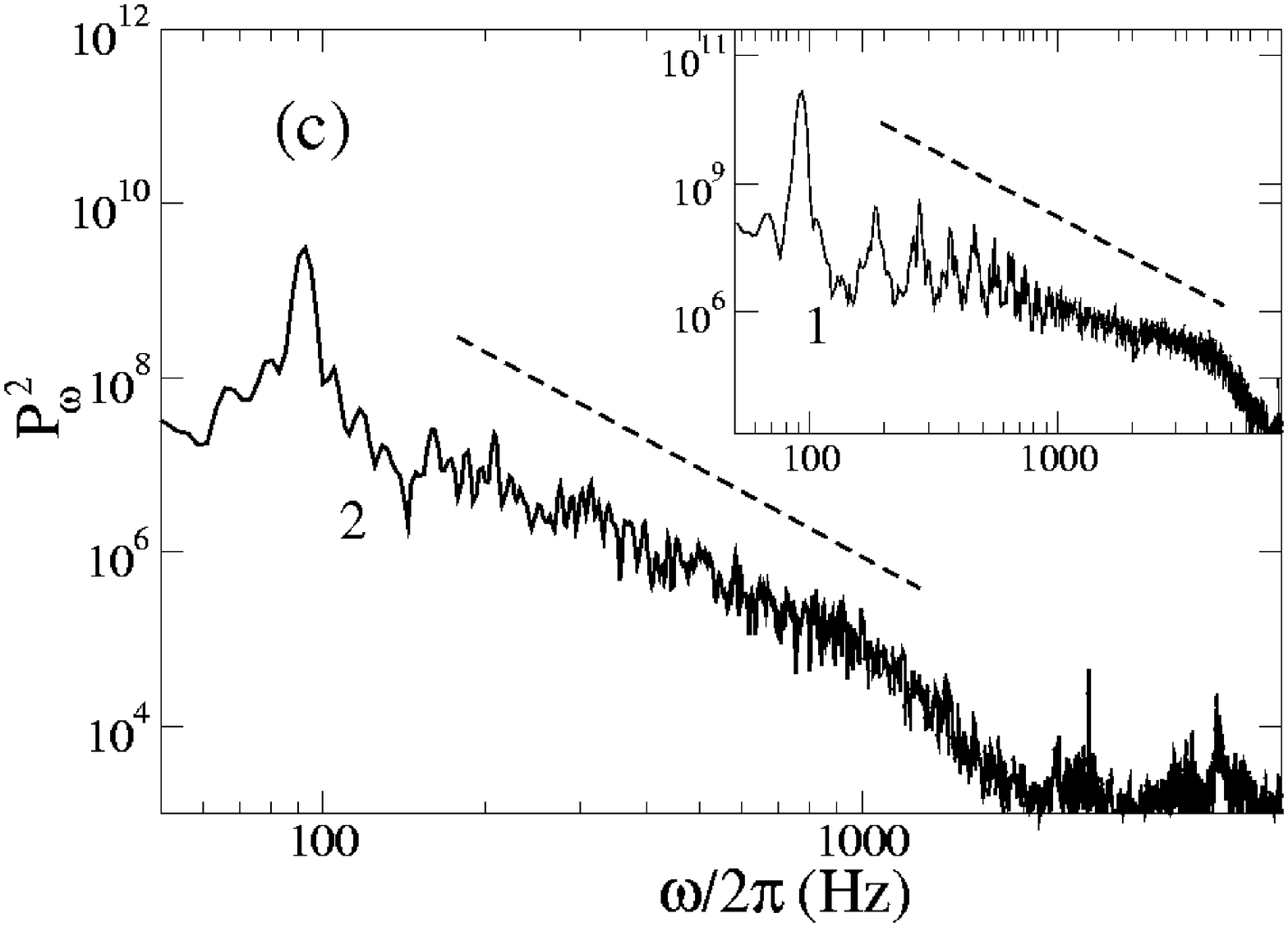}
\caption{\label{fig1} (a) The measured signal $P(t)$. The periodic
driving force was switched off at time $t=1.8$ s. (b) Evolution of
the turbulent power spectrum during the decay, calculated over
$P(t)$. Grey shading indicates frequency components in the power
spectrum whose $P_{\omega}^2$ exceeds the threshold $10^4$ (a.u.)
in the graph below. (c) Instantaneous power spectra calculated at
times indicated by the arrows in (b): curve 1 at the insert
corresponds to time $t=2$ s; curve 2 corresponds to $t=2.5$ s. The
spectra shown represent an ensemble average over 10 identical
measurements. The dashed  line in (c) corresponds to the
power-like dependence $P_{\omega}^2 \sim \omega^{-7/2}$ predicted
by theory \cite{Filonenko,Falkovich}.}
\end{figure}


{\it Experimental observations.} The experimental arrangements
were similar to those used in our earlier studies of steady-state
turbulence on the charged surface of liquid hydrogen~\cite{PTE}.
The measurements were made using an optical cell inside a helium
cryostat. Hydrogen was condensed into a cup formed by a bottom
capacitor plate and a guard ring 60 mm in diameter and 6 mm high.
The layer of liquid was 6 mm thick. The top capacitor plate (a
collector 60 mm in diameter) was located at a distance of 4 mm
above the surface of the liquid. A two-dimensional positive charge
layer was created just below the surface of the liquid with the
aid of a radioactive plate placed at the bottom of the cup. The
temperature of the liquid was 15--16 K. The waves on the charged
surface were excited by a periodic driving voltage applied between
the guard ring and the upper electrode. They were detected from
the variation of the total power $P(t)$ of a laser beam reflected
from the oscillating surface, which was measured with a
photodetector, sampled with an analogue-to-digital converter, and
stored in a computer. Given the size of the light spot, the
correlation function $I_{\omega}=\langle |\eta_{\omega} |^2
\rangle$ of the surface elevation $\eta({\bf r},t)$ in frequency
representation is directly proportional to the squared modulus of
the Fourier transform of the detected signal, $I_{\omega} = const
\, P_{\omega}^2$ at frequencies above 50 Hz.

In systems of finite size, the power spectrum of capillary
turbulence is discrete; but, at frequencies much higher than the
lowest resonant frequency (which in our experiments was $\sim$3
Hz), the intrinsic spectrum of resonant frequencies becomes
quasi-continuous due to viscous broadening of the resonances
and/or nonlinear broadening at pumping rates exceeding a critical
value (see theory \cite{Push,Pisma} and observations
\cite{Putterman,Alstrom,Lev1,LevEPL}). To establish the steady
turbulent state at the surface of the liquid, a 95 Hz ac driving
voltage was applied for $\sim$10 s. It was then switched off, and
we observed the relaxation oscillations of the surface. The
instantaneous power spectrum $P_{\omega}^2$ of nonstationary
surface oscillations was calculated by using a short-time Fourier
transform of the measured signal $P(t)$ \cite{Mallat}.

It is clearly evident from Fig.\ 1 that, during decay of the
turbulence, it is the {\it high frequency} components of the power
spectrum that are damped first. The amplitude of the main peak at
the driving frequency remains larger than the amplitudes of peaks
at the harmonics, all the time, both before and after the driving
force is switching off, i.e.\ the surface of the liquid continues
to oscillate mainly at the driving frequency. The wave amplitude
in the turbulent distribution decreases homogeneously during the
decay, and the power-law dependence of the spectrum persists, even
for low frequencies down to 100 Hz.

{\it Numerical calculations.} To try to understand the peculiar
form of decay, we modelled numerically the time evolution of the
initial steady-state turbulent power spectrum of the capillary
waves. According to WT theory \cite{Book,Filonenko} the evolution
of a turbulent spectrum can be described by the
integro-differential kinetic equation (KE) for the classical
occupation numbers $n_k$ of surface waves, where $k$ is a wave
vector. In our calculations we used the so-called ``local
approximation" for the KE, where the collision integral was
approximated by a differential operator. This approach is similar
to the well-known Fokker-Plank equation approach in the kinetic
theory of gases \cite{Landau}, and is based on nonlinear
interactions between waves of comparable frequencies. Such a
differential analogue of the KE can be written in a unique way
based on the known symmetry and scaling properties of the
turbulent state. A similar phenomenological model of WT has been
used successfully before, e.g.\ in studies of gravitational waves
on the surface of liquid \cite{Newell} and for optical turbulence
\cite{Opt}, yielding results that are in a good agreement with the
``exact" WT theory. To accord more accurately with the real
conditions of the experiment, we introduce into (1) a viscous
damping term for waves at all frequencies, in addition to the
nonlinear term. Capillary turbulence has not hitherto been studied
using this approach.

Following the general recommendations of Refs. \cite{Newell,Opt}
we write the corresponding equation for the envelope curve of the
turbulent spectrum as:
\begin{equation}
{\partial n_{\omega} \over
\partial t} = {C \over \omega ^{4/3}} {\partial \over \partial \omega}
\left[ \omega^{7} n_{\omega} {\partial \over \partial \omega}
\left( \omega n_{\omega} \right) \right] - 2 \gamma_\omega
n_{\omega} + f_{\omega}(t) . \label{maineq}
\end{equation}
For convenience in comparing numerical results with our
experimental data, we use the occupation number in
$\omega$-representation, $n_{\omega} (t) = n_k(t) |_{k=k(\omega)}$
as the main characteristic of the turbulent distribution, where
$k=k(\omega)$ is the function inverse to the dispersion law for
the capillary waves $\omega=\omega_k$. The first term on the right
hand side of (\ref{maineq}) plays the role of the collision
integral in the KE, $\gamma_{\omega} \propto \omega^{4/3}$ is the
damping coefficient of capillary waves, the term $f_{\omega}(t)$
models the direct action of the driving force, and $C$ is a
dimensional constant. For capillary waves the correlation function
$I_{\omega} = {\rm const} \, n_{\omega}$ has the same scaling
properties as $n_{\omega}$.
\begin{figure}[th]
\includegraphics[width=65.mm]{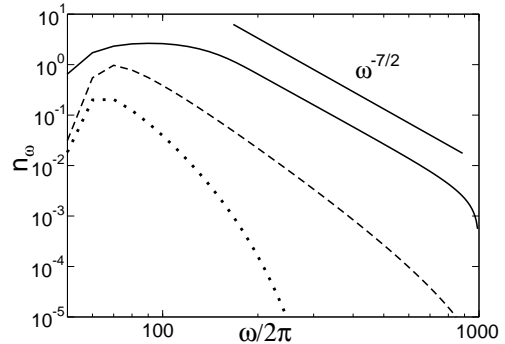}
\caption{\label{fig3} Decay of the turbulent spectrum. The
occupation number $n_{\omega}$ as a  function of renormalised
frequency calculated using (\ref{maineq}) at three renormalised
times: $t=0$, the initial steady-state spectrum (full curve);
$t=3$ (dashes); $t=10$ (dots). The straight line corresponds to
$n_{\omega} \sim \omega ^{-7/2}$ as predicted by the theory of
capillary turbulence \cite{Filonenko,Falkovich}. }
\end{figure}
In our simulations we renormalised time by a dimensional constant,
$t \rightarrow t/C$, and renormalised frequency, $\omega
\rightarrow C \omega $. We have assumed that the renormalised
driving frequency $\omega_d /2 \pi = 100$ to facilitate comparison
of our numerical modelling results for the free decay shown in
Fig.\ 2 with the experimental data of Fig.\ 1. The full curve
shows the envelope of the initial steady-state turbulent
distribution $n_{\omega}$. For the frequency range $200<\omega /2
\pi < 800$ it can evidently be approximated by a power-like
dependence $n_{\omega} \sim \omega ^{-7/2}$, in good agreement
both with the existing theory of capillary turbulence
\cite{Filonenko,Falkovich} and with our current and previous
experimental observations \cite{ZHETF}. The cut-off in the power
spectrum at the frequency $\omega_b /2 \pi \approx 800$ is caused
by viscosity. The dashed and dotted curves show the evolution of
the envelope after the driving force was removed. The destruction
of the power-like dependence clearly begins from the high
frequency domain of the turbulent spectrum. The energy of the
oscillations is concentrated mainly in the low frequency spectral
domain. That there is no destructive front propagating from low to
high frequencies is entirely consistent with our experimental
results.

These results may be explained qualitatively in terms of the
general theory of WT if we assume a fast redistribution of energy
between different frequency scales inside the inertial range,
leading to suppression of relaxation processes. Such a
redistribution will stabilize the power-like dependence of the
turbulent spectrum at low frequencies. A similar evolution of the
turbulent spectrum was observed in \cite{LevBound} with a smooth
decrease of driving amplitude, where the turbulent system remained
continuously in its steady state. Based on these observations we
can claim that the evolution of the freely decaying turbulent
spectrum has a quasi-adiabatic character. The kinetic time of
nonlinear wave interactions $\tau_{k}(\omega)$ plays the role of a
fast time; the time $\tau_{v}(\omega)$, characterizing viscous
damping of the waves, plays the role of a slow time within the
inertial frequency range. From (\ref{maineq}) one can estimate the
frequency dependences of $\tau_{k}(\omega) \propto \omega^{-7/6}$
and $\tau_{v}(\omega) \propto \omega^{-4/3}$, and of their ratio
as $r(\omega) =\tau_{k} (\omega) / \tau_v (\omega) \propto (\omega
/\omega_b) ^{1/6} $. It was shown in \cite{LevBound} that the
inertial range is limited at high frequencies by the condition
$r(\omega_b)\sim 1$, and that inside the inertial range the
parameter $r$ is small, i.e.\  nonlinear wave interactions occur
faster than the viscous damping processes. At these frequencies
the ratio $r(\omega)$ can be considered as a small adiabatic
parameter. At frequencies of the order of $\omega_b$, $r(\omega_b)
\sim 1$, so that the nonlinear and viscous processes have
comparable rates, and the adiabatic condition is violated. During
the decay process the amplitudes of the waves decrease, so the
kinetic time decreases as well \cite{ZHETF,LevBound}. Hence the
shift of the boundary frequency $\omega_b$ towards the low
frequency domain. At large times after the beginning of the decay,
$\omega_b$ becomes comparable with the driving frequency
$\omega_d$. Dissipation is then starting to play an important role
even at low frequencies. Thus we can claim that, for the
understanding of nonstationary turbulent processes, viscous losses
in a turbulent system are of central importance at all frequencies
above the driving frequency: we have seen, both in experiments and
in computations, that finite damping of the waves changes
qualitatively the character of the turbulent decay, compared to
the scenario envisaged earlier. Note that these considerations do
{\it not} apply to stationary turbulent phenomena, where
dissipation can be neglected completely over a wide range of
frequencies~\cite{Push,Zakharov2002}.

{\it Conclusion.} Our experimental observations and numerical
calculations have shown that the decay of capillary turbulence on
the surface of a normal liquid begins with the damping of high
frequency waves, so that the high frequency part of the power-like
turbulent spectrum is destroyed first. The energy-containing range
of frequencies does not shift toward the high frequencies in
practice, but remains at frequencies of the order of the driving
frequency. This remains true at all times during the decay, even
at large times when the turbulent regime is replaced by the
viscous damping of the  waves. Transition processes in the
turbulent system are damped by the fast nonlinear redistribution
of energy between waves whose frequencies are inside the inertial
range. These general conclusions may be useful for future studies
of nonstationary turbulent processes in a wide class of problems,
e.g.\ in hydrodynamics,  nonlinear sound in solids and liquids,
astrophysics, and plasma physics. In particular, our results
support the phenomenological propositions made in \cite{Ladik} to
account for the observed long-time properties of vorticity in the
decay of quantum turbulence in superfluid helium. It could be also
useful in relation to recent studies \cite{Pisma} of damping of
the monochromatic capillary wave on a liquid surface.

We acknowledge valuable discussions with V.E. Zakharov, E.A.
Kuznetsov and M.T. Levinsen, and thank V.N. Khlopinskii for
experimental assistance. The work was supported by the Leverhulme
Trust and the Engineering and Physical Sciences Research Council
(UK), INTAS (grant 2001-0618), by the Ministry of Industry,
Science and Technology of the Russian Federation (Contract No.\
40.012.1.1.11.64) and by the Russian Foundation of Basic
Researches (grant 03-02-16865). G.V.K. acknowledges support from
the Science Support Foundation (Russia).


\end{document}